\documentstyle{article}

\begin{document}

\renewcommand{\thefootnote}{\alph{footnote}}

\begin{titlepage}

\begin{flushleft}
TU-543  
\end{flushleft}
\
\\
\
\\

\begin{center}
{\LARGE \bf
Thermal Hair of Quantum Black Hole  
}
\end{center}

\ \\
\ \\

\begin{center}
\Large{
Y. Itoh\footnotemark[1],
M. Hotta\footnotemark[2],  
T. Futamase\footnotemark[3] and 
M. Morikawa\footnotemark[4]  
 }\\

{\it 
${\ }^{a, c}$ Astronomical Institute, Faculty of Science, Tohoku University, 
\\ Sendai 980-77, Japan \\
${\ }^{b}$ Department of Physics, Faculty of Science, Tohoku University,\\
Sendai 980-77, Japan\\
${\ }^{d}$ Department of Physics, Faculty of Science, \\ 
Ochanomizu Women's  University, Tokyo, Japan\\ 
}
\end{center} 
\
\\

\begin{abstract}

We investigate the possibility of statistical explanation of 
the black hole entropy by counting quasi-bounded modes 
of thermal fluctuation in two dimensional black hole spacetime. 
The black hole concerned is quantum in the sense that it is in thermal 
equilibrium with its Hawking radiation.  
It is shown that the fluctuation around such a black hole obeys 
a wave equation with a potential whose peaks are located near the black hole
and which is caused by quantum effect. 
We can construct models in which the potential in the above sense 
has several positive peaks and there are quai-bounded modes confined between 
these peaks.    
This suggests that these modes contribute to the black hole entropy.
However it is shown that the entropy associated with these modes dose not
obey the ordinary area law. Therefore we can call these modes as  
an additional thermal hair of the quantum black hole.

\end{abstract}

\footnotetext[1]{Email address: yousuke@astr.tohoku.ac.jp}
\footnotetext[2]{Email address: hotta@tuhep.phys.tohoku.ac.jp}
\footnotetext[3]{Email address: tof@astr.tohoku.ac.jp}
\footnotetext[4]{Email address: hiro@phys.ocha.ac.jp}

\end{titlepage}

\section{Introduction}

Since the discovery of Hawking radiation \cite{H}, there has been a 
growing interest on the close relation between gravity and thermodynamics. 
In particular, the black hole entropy problem, whether or not black hole 
entropy can be explained statistically has attracted much attention 
because it is hoped that the statistical explanation may shed light 
on the  way to quantize gravity. 
In fact recent development in string theoretical explanation for black
hole entropy \cite{SV} is regarded as the indication that string theory may be an integral ingredient of quantum gravity. 

Although these development excites us, it raises at the same time an uneasy 
feeling how can quantum gravity plays such 
an essential role in understanding entropy of a black hole. 
This is because the entropy is associated 
with any size of black hole whose gravity is very week.  
Also, we note that such an exact explanation is 
only successful so far for BPS black hole.  
Thus it seems us to be also important to think about other 
possibilities to explain entropy statistically from field theoretical 
view point. 

There has been various attempts in this direction, 
but all of them so far considered a classical black hole 
which is not in thermal equilibrium with radiation. 
For example York has attempted to solve this problem by counting 
quasi-normal modes in quantum ergo-region \cite{York}. 
Although the approach is very interesting, thermal nature of the 
black hole is put in by hand and some details are 
left unexplained partly because of the difficulty associated with 
the complexity in four dimensional gravity, namely there is no hope to 
have a realistic solution in some sort of theory of quantum gravity 
expressing a black hole in thermal equilibrium with radiation. 

There might be something important missing if we are not 
paying any special attention to the situation 
where a radiating black hole is in thermal equilibrium 
with incoming radiation. Therefore it seems urgent to examine carefully 
the entropy of a black hole in thermal equilibrium from field 
theoretical view point before withdrawing any conclusion. 
This is main motivation of this paper. 

Here we shall follow in spirit York's approach to the problem, 
but with two different settings. 
One is that we consider two-dimensional dilaton gravity 
because of its simplicity instead of Einstein gravity. 
We hope this will not much make the problem unrealistic 
because it has been shown that two-dimensional gravity 
shows many common feature with Einstein gravity on black hole formation 
and Hawking radiation \cite{RST,CGHS,L,NO}.  
Second is we consider the entropy of a quantum black
hole in the sense that it is in thermal equilibrium with radiation. 
As stressed above such setting seems more appropriate to think 
about thermodynamical quantities like entropy.   

Thus we consider thermal fluctuation 
around a quantum black hole in two dimensional dilaton gravity and 
if there are some kind of (quasi-) stable modes which 
could contribute to its entropy.

To avoid any confusion, we make clear our situation here. 
We mean by a quantum black hole as a static configuration 
in a pure state defined by an effective action including 
quantum effect. Then the quantum black hole is supposed to be in 
equilibrium with a thermal bath. Thus we concern the thermal state and
the thermal fluctuation will be inevitably present. 
In our paper the word "fluctuation" is always used in this sense.

Actually the above program has been carried out by 
the present authors for investigating the dynamical 
instability of the quantum black hole in Lowe's model \cite{IHMF}.
There it is found that the fluctuation obeys the wave equation 
with a negative definite potential and thus such a black hole is 
unstable. This does not mean that quantum black holes in any two-dimensional models are unstable. 
This is because the dilaton gravity in two dimensions does not 
have a unique action and there exist an infinite number of variations 
of the models. 
Thus it may well be other models in which the potential for the fluctuation 
behaves totally different such that quasi-stable modes exit between 
its positive peaks.  What we would like to show is that 
this is actually the case. For this purpose we need 
a general model of two-dimensional dilaton gravity which admit a family of 
static black hole solution with regular horizon. 

Fortunately for classical dilaton gravity models this is provided 
by some of the present authors \cite{FHI}. 
They have given a prescription to construct all such dilaton gravity 
models containing at most second derivatives of the fields 
in the action. We shall extend their formalism into quantum models and 
carry out the above program.

This paper is organized as follows.  In section 2 we shall give a brief
review of our general prescription how to construct  a quantum black
hole solution in general two-dimensional dilaton gravity.  
Here the previous studies play  essential roles \cite{FHI}. 
Some details of calculation are found in appendix A.  
In section 3 the equation for the fluctuation is obtained by perturbing 
the solution obtained in the previous section. Calculational details are 
found in Appendices B and C. 
In section 4 we shall propose new type models in which quasi-bounded modes
are present. 
Namely we will find a model in which the potential for 
the fluctuation has double peaks.  
Some discussions are devoted in section 5 where  we shall
also summarize our results.  
In appendix D  we will apply the present formalism  
to the problem of dynamical instability of a black hole \cite{IHMF}.  
There the solvable models are discussed in some detail. 
In particular we will find RST model \cite{RST} is dynamically stable.

\section{Quantum black hole in general two-dimensional dilaton gravity}
\ \\
   In this section we shall give a brief review of how to construct 
a quantum black hole solution in general two-dimensional dilaton gravity. 
More detailed discussion can be seen in the appendix A.

   We consider the following action which includes only second derivative
terms, the dilatonic cosmological term and back reaction term.
\begin{eqnarray}
S     &=&      
            \frac{1}{2\pi}\int d^2 x \sqrt{-g}
            \left(
                   F(\kappa,\phi) R
                   +4G(\kappa,\phi)(\nabla \phi)^2
                   +4\lambda^2 U(\kappa,\phi)
            \right)\nonumber\\
\mbox{} &-& \frac{\kappa}{8\pi} \int d^2 x \sqrt{-g} R\frac{1}{\nabla^2}R.
\label{action}
\end{eqnarray}
Here, $g$ is the determinant of the two-dimensional metric,  $\phi$ is 
the dilaton field.  $\kappa$ is 
related to the number of matter fields $N$, such as $\kappa = \frac{N}{12}$. 
The back reaction term is obtained from integrating out the
fluctuation in massless matter fields in the same way as the CGHS 
model \cite{CGHS}. 
The $F, G$ and $U$ are certain functionals of $\phi$ and specified later. 
This action without back reaction term is shown to be general in the
sense that it admit a flat vacuum solution and a family of static black
hole solutions with regular horizon \cite{FHI}.

We shall adopt the conformal gauge for the metric throughout this paper. 
\begin{eqnarray}
&&
x^{\pm} = t \pm r,\\
&&
ds^2 = -e^{2\rho}dx^{+}dx^{-} .
\end{eqnarray}
Here, $t$ and $r$ is timelike and spacelike coordinate respectively.

Then the equations of motion and constraint equation are simply calculated 
by variating the action (\ref{action}) as follows.
\begin{eqnarray}
&&
2F' \partial_+ \partial_- \rho
+ 8G\partial_+ \partial_- \phi
+ 4G' \partial_+ \phi \partial_- \phi
+ \lambda^2 U' e^{2\rho}  = 0, \label{motion1} \\
&&
\partial_+ \partial_- F  
+ \kappa \partial_+ \partial_- \rho
+ \lambda^2 U e^{2\rho} = 0, \label{motion2}  
\end{eqnarray}
and 
\begin{eqnarray}
  \partial_{\pm}^2 F 
- 2\partial_{\pm} \rho\partial_{\pm}F
- 4G(\partial_{\pm} \phi )^2
= - \kappa [\partial_{\pm}^2 \rho  - (\partial_{\pm} \rho)^2 + t_{\pm}], 
\label{constraint}
\end{eqnarray}
where prime denotes the derivative with respect to $\phi$ and  
dot to $r$ in the following. 

For the static solutions, the equations of motion (\ref{motion1}),
(\ref{motion2}) are simplified as
\begin{eqnarray}
&&
  F' \ddot{\rho} 
+ 4G\ddot{\phi}
= 2\lambda^2 U' e^{2\rho} 
- 2G'\dot{\phi}^2, \label{static1} 
\\
&&
 \kappa \ddot{\rho} 
+ F'\ddot{\phi}
= 4\lambda^2 U e^{2\rho}
- F'' \dot{\phi}^2. \label{static2} 
\end{eqnarray}
Using these equations, we can write the constraint equation 
(\ref{constraint}) in the following form 
\begin{eqnarray}
 \lambda^2 U e^{2\rho}
= G\dot{\phi}^2
+ \frac{1}{2}F'\dot{\rho}\dot{\phi}
+ \frac{\kappa}{4}(\dot{\rho}^2 - \lambda^2). \label{static3}
\end{eqnarray}
$t_{\pm}$ is specified as $\lambda^2$ from the following boundary 
conditions which are necessary to have a static black hole with regular
horizon:
\begin{eqnarray}
\rho (r\sim -\infty )
                      &=& \lambda r + \rho_o + \rho_1 e^{2\lambda r} 
                      + \rho_2 e^{4\lambda r}+ \rho_3 e^{6\lambda r}
                      + \cdots,  
\label{rhosBC} \\ 
\phi(r\sim -\infty) &=& \phi_o + \phi_1 e^{2\lambda r} 
                     +  \phi_2 e^{4\lambda r} + \phi_3 e^{6\lambda r}
                     +  \cdots.
\end{eqnarray}
where $\rho_o$ etc. are some constant. 

Then we consider the condition for the existence of the flat vacuum
solution:
\begin{eqnarray}
\rho &=& const, \nonumber \\
\phi &=& \phi(r). \nonumber
\end{eqnarray}

By substituting these into the equations (\ref{static1}), 
(\ref{static2}) and (\ref{static3}) we find that $F, G$ and $U$ must
satisfy the condition: 
\begin{equation}
\frac{U'}{U} = \frac{G'}{G} -2\frac{F''}{F'} +8\frac{G}{F'}.
\label{flat}
\label{flatvacuum}
\end{equation}
In the following we assume that $F, G$ and $U$  obey this condition.

Now, for our purpose it turns out convenient for later discussions that we determine
$F$ from an arbitrary gravity field $\rho$ (see appendix A for detail).  
From equations (\ref{static1}), (\ref{static2}), 
(\ref{static3}) and (\ref{flatvacuum}), 
we obtain the following equation for $F$,   
\begin{eqnarray}
&&
\left[2\dot{\rho} 
+\kappa\frac{\dot{\rho}^2 -\lambda^2}{\dot{F}} \right]
\frac{1}{\dot{F}}F^{(3)}
\nonumber\\
&&
-\left[ 
4\dot{\rho} -2\kappa
\frac{\ddot{\rho} -\dot{\rho}^2 +\lambda^2}{\dot{F}}
\right]
\left( \frac{\ddot{F}}{\dot{F}} \right)^2
\nonumber\\
&&
-\left[
2\ddot{\rho}
-4\dot{\rho}^2
+6\kappa 
\frac{2\dot{\rho}\ddot{\rho} -\dot{\rho}(\dot{\rho}^2 -\lambda^2)}
{\dot{F}}
-2\kappa^2
\frac{\ddot{\rho}-\dot{\rho}^2+\lambda^2}{\dot{F}}
\frac{2\ddot{\rho}-\dot{\rho}^2 +\lambda^2}{\dot{F}}
\right]\frac{\ddot{F}}{\dot{F}}
\nonumber\\
&&
-2\kappa
\frac{\ddot{\rho}^2 -2\dot{\rho}^2 \ddot{\rho} 
-\dot{\rho}\rho^{(3)}  }
{\dot{F}}
-\kappa^2
\frac{8\dot{\rho}\ddot{\rho}^2 
-\left(\rho^{(3)} +6\dot{\rho}\ddot{\rho} \right)
(\dot{\rho}^2 -\lambda^2 ) }
{\dot{F}^2}
+2\kappa^3
\frac{\ddot{\rho}}{\dot{F}}
\left[ 
\frac{\ddot{\rho}-\dot{\rho}^2 +\lambda^2}{\dot{F}}
\right]^2
=0. \label{FsEQ}  
\end{eqnarray} 
This equation determines $F$ uniquely up to three integral constants
once one choose a certain gravity
field $\rho$ and the boundary condition at the horizon.

From the equation  (\ref{rhosBC}) and (\ref{FsEQ}), $F$ must have the form
at the horizon, 
\begin{eqnarray}
F(r\sim -\infty)
&=& F_o + F_1 e^{2\lambda r} + F_2 e^{4\lambda r}
+ F_3 (F_1 ,F_2 ,\rho_1 ,\rho_2 ,\rho_3 ) e^{6\lambda r} 
+\cdots .
\label{FsBC}
\end{eqnarray}
Thus the free boundary conditions for $F$ are $F_1$ and $F_2$. 
Once the boundary conditions $F_1 ,F_2 ,\rho_1 ,\rho_2$ and $\rho_3$ are
given, we can solve the equation (\ref{FsEQ}) numerically to obtain 
the static solution expressing a quantum black hole (see detail in
Appendix A). 
$F_o$ plays no physical role, but it may chosen to make  $F \geq 0$
because the inverse of $F$ can be interpreted as the gravitational constant
(see the action (\ref{action})).

\section{Linear Perturbation around a Quantum black hole}
\ \\
In this section we will discuss how to write down the perturbation
equation.
Now suppose we solved the equation (\ref{FsEQ}) numerically to have a
static solution which expresses a quantum black hole in equilibrium with 
a thermal bath. Then we shall perturb the solution to obtain the equation
for the fluctuation. This can be done as follows (details are found in
Appendix B). 
First, we split our variables $\phi$ and $\rho$ as 
\begin{eqnarray}
    \phi &=& \phi_{b}(r) + \delta \phi(x),   \label{deltaP}\\
    \rho &=& \rho_{b}(r) + \delta \rho(x),   \label{deltaR}
\end{eqnarray}
where $ \phi_{b}(r)$ and $ \rho_{b}(r) $ are  static background solutions
and obey equations (\ref{static1}), (\ref{static2}) and 
(\ref{static3}). We assume the perturbed variables $\delta \phi$ and 
$\delta \rho$ to be small in magnitudes.
Perturbing the equations of motion (\ref{motion1}), (\ref{motion2})
and (\ref{constraint}) and with the aid of the conformal invariant variable:
\begin{eqnarray}
\Phi &=& \delta \rho - \frac{\partial}{\partial r} 
                 \left[\frac{\delta \phi}{\dot{{\phi}_b}}\right] 
               - \frac{d\rho}{d\phi}|_b \delta \phi,  
\end{eqnarray} 
we can write down the equation of the thermal fluctuation $\Phi$ as  
\begin{eqnarray}
    \left[  \frac{\partial^{2}}{\partial t^{2}} 
      - \frac{\partial^{2}}{\partial r^2}
      + V[\rho_{b}(r),\phi_{b}(r)] \right] \tilde{\Phi}(x)
     &=& 0 \label{wave}, 
\end{eqnarray} 
where we define a new variable $\tilde{\Phi}$ by
\begin{eqnarray}
    \Phi &=& 
         \exp \left[2 \int \frac{\Omega}{\dot{{\phi}_b}} dr \right]
         \tilde{\Phi} \label{phitilde}
\end{eqnarray}
with
\begin{eqnarray}
    \Omega &=& 
             - 2 \lambda^{2} e^{2 \rho_b}
               \frac{F'U - \frac{\kappa}{2} U'}{F'^2 - 4 \kappa G}.
\end{eqnarray}
The explicit form of the potential $V$ is
\begin{eqnarray}
V&=&
\frac{\kappa}{2[
\dot{\tilde{F}}^2 -\kappa\ddot{\tilde{F}}-\kappa^2 \lambda^2
 ]^2}
\nonumber\\
&&\times\left[
\left( 
 \tilde{F}^{(4)} -4\lambda^2 \ddot{\tilde{F}}
\right)
(\dot{\tilde{F}}^2 -\kappa \ddot{\tilde{F}}-\kappa^2 \lambda^2 )
\right. \nonumber\\
&&\left. \ \ \ 
+6\left(
\ddot{\tilde{F}}^3 -\dot{\tilde{F}}\ddot{\tilde{F}} \tilde{F}^{(3)}
+\frac{\kappa}{4}
\left[
\left( \tilde{F}^{(3)} \right)^2
+4\lambda^2 \ddot{\tilde{F}}^2
\right]
\right)
\right],
\label{POTinF}
\end{eqnarray}
where 
$\tilde{F}$ is defined by
\begin{eqnarray}
\tilde{F}=F+\kappa \rho_b  \label{Ftilde}
\end{eqnarray}
and $\tilde{F}^{(n)}$ means n-th derivative of $\tilde{F}^{(n)}$. 
 
To make discussion clear we shall present  
an expression for the above potential to the leading order of
$\kappa$ (see appendix C).
Although the back reaction term in the action (\ref{action})
is derived in  the large $\kappa$-limit, such an expression for the
potential make sense because the term  proportional to $\kappa$ is relatively 
small when mass of black hole become large. 
Furthermore it is well worth noting that the shape of the potential 
(\ref{POTinF}) calculated to 
full order of $\kappa$ looks like that calculated to leading order of
$\kappa$. Also to obtain the potential to leading order 
of $\kappa$, one must know only the classical background solution.

The classical solution of the equation (\ref{FsEQ}), $F_{c}$ is   
\begin{eqnarray}
\dot{F}_c = \frac{2 e^{- \rho_c(\infty)} \mu}
                 {e^{-2 \rho_c} -e^{-2 \rho_c(\infty)}},
\label{classicalF}
\end{eqnarray}
where $\mu$ is a black hole mass and $\rho_{c}$ denotes the classical 
back ground gravity field.
Then the potential  $V_{c1}$ to lowest order of $\kappa$   becomes
\begin{eqnarray}
V_{c1}
&=&
\frac{\kappa}{2\dot{F}^4_{c}}
[\dot{F}^2_{c} F^{(4)}_{c} -4\lambda^2 \dot{F}^2_{c} \ddot{F}_{c}
+6\ddot{F}^3_{c} -6\dot{F}_{c} \ddot{F}_{c} F^{(3)}_{c} ] \nonumber \\
\mbox{} &=&
\frac{\kappa e^{\rho_c(\infty)-2\rho_c}}{2 \mu}
\left[
\rho^{(3)}_{c}
-6\dot{\rho}_{c} \ddot{\rho}_{c} +4\dot{\rho}_{c}^3
-4\lambda^2 \dot{\rho}_{c} 
\right]. \label{perturbPOT}
\end{eqnarray}

Furthermore, we can obtain the formula of a potential for black hole with 
larger mass $M > \mu$. First we introduce a new spacelike coordinate $r'$ 
to which spacetime of a black hole with mass $M$ is referred;
\begin{eqnarray}
r'&=&
\frac{M}{\lambda \mu} \dot{\rho_c}(r_H)
\int\frac{1}{\frac{M}{\mu}+(1-\frac{M}{\mu}) 
e^{2\rho_c(\infty)-2\rho_c(r)} } dr.
\label{integratedr'inr}
\end{eqnarray} 
Then the horizon of a mass $M$ black hole in $r'$ coordinate is 
located at $r_H$ in $r$ coordinate,  defined by
\begin{eqnarray}
&&
e^{2\rho_c(r_H)} = e^{2\rho_c(\infty)} \left[1-\frac{\mu}{M} \right] 
\label{rH}
\end{eqnarray}
At last, we may have the potential ${V'}_{c1}(r')$ for a mass $M$ black hole
\begin{eqnarray}
{V'}_{c1}(r')
&=&\frac{\kappa\lambda^2 e^{-\rho_c(\infty)}}{2 \dot{\rho_c}(r_H)^2 }
\frac{\mu e^{2\rho_c(\infty)-2\rho_c(r)}}{M^2}
\left[\frac{M}{\mu} +(1-\frac{M}{\mu})e^{2\rho_c(\infty)-2\rho_c(r)} \right]
\nonumber\\
&&\times
\left[
\frac{d^3 \rho_c(r)}{dr^3}
-\frac{6 \frac{M}{\mu} + 2(1-\frac{M}{\mu})e^{2\rho_c(\infty)-2\rho_c(r)}}
      {\frac{M}{\mu} +(1-\frac{M}{\mu})e^{2\rho_c(\infty)-2\rho_c(r)}}
\frac{d\rho_c(r)}{dr}\frac{d^2 \rho_c(r)}{dr^2}
\right. \nonumber\\
&&\left. \ \ \ 
+
4 \frac{M^2}{\mu^2} 
\frac{d\rho_c(r)}{dr}
\frac{\left(\frac{d\rho_c(r)}{dr} \right)^2 -\dot{\rho_c}(r_H)^2}
{\left[\frac{M}{\mu} +(1-\frac{M}{\mu}) e^{2\rho_c(\infty)-2\rho_c(r)}
 \right]^2}
\right].
\label{POTinLarge} 
\end{eqnarray}

This expression will be used in the next section.

\section{New Model} 
In this section we propose new models in which quasi-bounded modes are found. 

The basic idea is as follows. There may be a large
region of $(F,G,U)$ space where the model which is specified by $F,G$
and $U$ has positive potential  defined 
by (\ref{POTinF}) with some peaks. These peaks are located 
at a distance of about $\lambda$ from the black hole.     
The potential is made by the fluctuations of matter fields 
which are inevitably present in thermal equilibrium through 
the quantum back reaction. 
Then the thermal fluctuation of the radiation may be confined into 
the regions surrounded by these peaks. 
Such quasi-trapped modes are neither growing nor decaying when black hole is
in thermal equilibrium. 
Thus it can be seen from the distant observer as if the black hole may be surrounded by these modes and we can think that they contribute to 
the black hole entropy.    

Now, we show one of such new models. For simplicity, we will give 
a model which has two potential peaks. 
We note that there are models in which the potential 
has many peaks and such models are not rare.

First we take black hole field such as
\begin{eqnarray}
\rho (r) &=& - 0.032 - \frac{1}{2} \ln (1 + e^{-2 \lambda (r-1)})
\nonumber \\
\mbox{}  &+&   \frac{0.46}{\cosh [2 \lambda (r-0.5)]} 
          + 0.032 \frac{e^{2 \lambda (r+1.5)}}{1+e^{2 \lambda (r+1.5)}}.
\label{rho}             
\end{eqnarray}
(see Figure 1.)
One can easily confirm that this satisfies the condition (\ref{rhosBC}).

Then from the equation (\ref{FsEQ}), we can solve $F$ numerically and
which is positive definite. 
We show $\dot{F}$ in Figure 2. This shows that the model
which has above black hole field really exists.   
The potential can then be calculated numerically by the equation
(\ref{POTinF}) and is shown in Figure 3.
This potential has in fact two peaks. From this potential, we can find
the quasi-bounded modes by solving the equation (\ref{wave}) numerically.

Now let's discuss the relation between the black hole mass $M$ and the
number of quasi-bounded modes.
Here we consider the potential to the leading order of $\kappa$ 
(see Figure 4).
To estimate the number of modes confined between the peaks, 
we consider the area between these peaks.   
The equation (\ref{POTinLarge}) may be used to 
examine the behavior of the potential in black hole with various mass $M$. 
Here we consider two cases. 
In one case the cosmological term $\lambda$ retain its value 
when $M$ increase (Figure 5).
In the other $\lambda$ scales as $M^{-1}$ which is expected in a four
dimensional black hole case (Figure 6).
It can be seen that in both cases  the horizon 
erode away the potential more and more as $M$ becomes larger. 
For example, in Figure 5 the left hand side potential peak becomes 
smaller in height
when $M$ becomes larger, and beyond a certain value of mass 
 the peak disappears. Then the 
 (quasi-) bounded modes confined between these peaks must vanish.
Thus the area between the peaks of the potential tends to decrease 
when $M$ is increased.

Since the entropy is expected to increase as the mass of black hole, 
this behavior suggests that the freedoms associated with 
the quasi-bounded modes may not contribute to the ordinary black hole entropy.  
We might call them an additional thermal hair. 
It is very interesting to see if these thermal hair has something to do with 
information paradox in black hole evaporation \cite{S}.

\section{Summary}
In this section we devote summary and some discussions.

We have investigated  the possibility of explaining the black hole entropy 
statistically by quasi-bounded modes of thermal fluctuation of black 
hole space time. 
We consider the black hole in thermal equilibrium with the 
Hawking radiation and it is natural situation to think about black hole 
entropy. We have found the following result. 
The fluctuation obeys wave equation with a
potential which is near the black hole  and is caused by quantum effect. 
There are many models in which the potential has
positive peaks and quai-bounded modes exist between these peaks. These
modes are confined near the black hole. Thus    
they may contribute to the black hole entropy. 
But ordinary area law is not found, therefore we conclude 
that the entropy related with these modes is not ordinary black hole entropy.  
We can identify them as an additional thermal hair.
Our result suggests that the black hole entropy problem is not 
settled down at all and actually the situation becomes more difficult. 
  
Finally, we note that there is a possibility to have 
ordinary entropy from quasi-bound modes discussed above. 
We have shown that the potential is eroded by the horizon when black hole
mass becomes large in the previous section. 
This is understood by the behavior of $\rho$ and the equation (\ref{rH}). 

However, there is a difference between two cases, 
in one case $\lambda$ is not scaled with  mass, and 
in the other case $\lambda$ is scaled with mass. 
In the latter case the potential
is more stretched in width than in the former case.  
Thus we may have a model in which   
there are some potential peaks which are 
stretched in width and are not eroded by the horizon 
even if the mass becomes large. 
Then the number of the quasi-bounded modes increase 
as black hole mass. This will remain as a future work.

\appendix
\section{Inverse Approach to  Black Hole Solution}

In this section we will discuss the way to determine $F$ from gravity field
$\rho$ up to integral constants and derive the equation (\ref{FsEQ}). 
This point of view is found in \cite{FHI} for classical
two dimensional dilaton gravity models. Here we discuss its quantum
version.  

First,  we consider {\it functionals} $F,G$ and $U$ of $\phi$ as 
{\it functions} of $r$.

\begin{eqnarray*} 
F(\kappa, r) &=& F(\kappa, \phi (r) ),
\\                               
\bar{G}(\kappa, r) &=& \dot{\phi}^2 G(\kappa, \phi (r)),
\\                                  
U(\kappa, r) &=& U(\kappa, \phi (r)),
\end{eqnarray*}
where we introduce $\bar{G}$.  
Then the equations of motion (\ref{static1}), (\ref{static2}) and 
constraint equation (\ref{static3}) can be rewritten as
\begin{eqnarray}
&&
\dot{F} \ddot{\rho}  =
2\lambda^2 \dot{U} e^{2\rho} -2\dot{\bar{G}},
\label{static1.1} \\
&&
\kappa \ddot{\rho} =4\lambda^2 U e^{2\rho}
-\ddot{F}, \label{static2.1} \\
&&
\lambda^2 U e^{2\rho}
=
\bar{G}
+
\frac{1}{2}\dot{F}\dot{\rho}
+
\frac{\kappa}{4}(\dot{\rho}^2 - V_o ).
\label{static3.1}
\end{eqnarray}
Here we have changed the derivative with respect to 
$\phi$  by  the derivative with respect to $r$ as follows.
\begin{eqnarray}
&&
F' = \frac{1}{\dot{\phi}} \dot{F},
\nonumber\\
&&
F'' =\frac{1}{\dot{\phi}^2 } \ddot{F} 
-\frac{\ddot{\phi}}{\dot{\phi}^3} \dot{F}
\nonumber
\end{eqnarray}
and so on.
Also,  the condition for existence of flat vacuum (\ref{flat}) 
becomes 
\begin{eqnarray}
2 \frac{\ddot{F}}{\dot{F}}
+\frac{\dot{U}}{U} -\frac{\dot{\bar{G}}}{\bar{G}}
=8\frac{\bar{G}}{\dot{F}}.
\label{condition1.1}
\end{eqnarray}
We note that $\phi$ is removed in the above equations
(\ref{static1.1}), (\ref{static2.1}), (\ref{static3.1}) and
(\ref{condition1.1}).

From the equations (\ref{static1.1}),
(\ref{static2.1}) and (\ref{static3.1}) we can solve for $G$ and $U$ as 
\begin{eqnarray}
&&
\bar{G}
=\frac{1}{4}
\left[
\ddot{F}-2\dot{\rho}\dot{F}
+\kappa(\ddot{\rho} -\dot{\rho}^2 + V_o )
\right], \label{GsEQ}
\\ 
&&
U
=\frac{e^{-2\rho} }{4\lambda^2}
\left[
\ddot{F}+\kappa \ddot{\rho}
\right]. \label{UsEQ}
\end{eqnarray}

By substituting these into (\ref{condition1.1}), we arrive at 
the equation
(\ref{FsEQ})

\section{Perturbations around static solutions} 

In this section we will study a general treatment of the linear
perturbation around the static background solutions and derive the
equations (\ref{wave}) and (\ref{POTinF}).  
First, we split our variables as in equations (\ref{deltaP}) and 
(\ref{deltaR}). 
Then the equations of motion for $\delta
\rho$ and $\delta \phi$ can be derived  from (\ref{motion1}) 
and (\ref{motion2}), respectively.
\begin{eqnarray}
&&
2 F' \partial_+ \partial_- \delta \rho
- \frac{1}{2} F''  \ddot{\rho}_b \delta \phi
+ 8 G \partial_+ \partial_- \delta \phi
- 2 G'  \ddot{\phi}_b \delta \phi
- G'' {\dot{\phi}_b}^2 \delta \phi
- 2 G' \dot{\phi}_b \delta \dot{\phi}  \nonumber \\
&&
+ \lambda^2 U''  e^{2 \rho_b} \delta \phi
+ 2 \lambda^2 U' e^{2 \rho_b} \delta \rho = 0,  \label{pertub1}\\
&&
F' \partial_+ \partial_- \delta \phi
- \frac{1}{2} F'' \dot{\phi}_b \delta \dot{\phi}
- \frac{1}{4} F''' {\dot{\phi}_b}^2 \delta \phi
- \frac{1}{4} F'' \ddot{\phi}_b \delta \phi
+ \kappa  \partial_+ \partial_- \delta \rho  \nonumber \\
&&
+ \lambda^2 U' e^{2 \rho_b} \delta \phi
+ 2 \lambda^2 U e^{2 \rho_b} \delta \rho  = 0. \label{perturb2}
\end{eqnarray}
Moreover, the linear perturbation of $<T_{\pm\pm}>$, the trace anomaly
part of the stress energy tensor: 
\begin{eqnarray*}
<T_{\pm\pm}>      &=&
        \kappa[  (\partial_{\pm}\rho)^{2} 
       - \partial^2_{\pm}\rho
       - t_{\pm}(x^{\pm})], 
\end{eqnarray*}
is calculated from the constraint equation (\ref{constraint}),
\begin{eqnarray}
\delta <T_{\pm\pm}> &=& 
\delta (\partial_{\pm}^2 F 
- 2\partial_{\pm} \rho\partial_{\pm}F
- 4G(\partial_{\pm} \phi )^2) \nonumber \\
\mbox{}  &=&  
F' {\partial_{\pm}}^2 \delta \phi
\pm F'' \dot{\phi}_b \partial_{\pm} \delta \phi
+ \frac{1}{4} F''' {\dot{\phi}_b}^2 \delta \phi
+ \frac{1}{4} F'' \ddot{\phi}_b \delta \phi  \nonumber \\
\mbox{}  &\mp&  
F' \dot{\phi}_b \partial_{\pm} \delta \rho
\mp F' \dot{\rho}_b \partial_{\pm} \delta \phi 
- \frac{1}{2} F'' \dot{\rho}_b \dot{\phi}_b \delta \phi 
- G' {\dot{\phi}_b}^2 \delta \phi
\mp 4 G \dot{\phi}_b \partial_{\pm} \delta \phi 
\end{eqnarray}

It is more convenient to use  
the conformal gauge invariant quantities defined below.  
\begin{eqnarray}
\Phi &=& \delta \rho - \frac{\partial}{\partial r} 
                 \left[\frac{\delta \phi}{\dot{{\phi}_b}}\right] 
               - \frac{d\rho}{d\phi}|_b \delta \phi,  \\
\Theta &=&  {\dot{\phi}_b}       
              \partial_{+}\partial_{-} 
              \left[\frac{\delta \phi}{\dot{{\phi}_b}} \right], \\
\Xi_{\pm}&=&  \delta <T_{\pm\pm}> 
            \mp 4\frac{<T_{\pm\pm}>_{b}}{\dot{{\phi}_b}}
            \partial_{\pm} \delta \phi \nonumber   \\
    \mbox{}     &+&   
            \left(  2\frac{\ddot{{\phi}_b}}{({\dot{\phi}_b)}^{2}} 
                    <T_{\pm\pm}>_{b} 
           - \frac{{<\dot{T}_{\pm\pm}>}_b}{{\dot{\phi}_b}} \right) 
                 \delta \phi, 
\end{eqnarray} 
where  $<T_{\pm\pm}>_b$ is the trace anomaly evaluated with the background 
solution $(\rho_b, \phi_b)$. 

We may obtain the equations for perturbed quantities with these
variables as 
\begin{eqnarray}
    0  &=& 
            2 F' 
           \left( 
                    \partial_{+}\partial_{-}\Phi 
                 +  \frac{\partial}{\partial r}
                    \left[\frac{\Theta}{\dot{{\phi}_b}}\right] 
                 +  \frac{\dot{{\rho}_b}}{\dot{{\phi}_b}} \Theta
           \right) 
        +   2 \lambda^2 U' e^{2 \rho} \Phi
        +   8 G \Theta,    \label{EQ1} \\
    0  &=& 
           \kappa 
           \left( 
                    \partial_{+}\partial_{-}\Phi 
                 +  \frac{\partial}{\partial r}
                    \left[\frac{\Theta}{\dot{{\phi}_b}}\right] 
                 +  \frac{\dot{{\rho}_b}}{\dot{{\phi}_b}} \Theta
           \right) 
        +   2 \lambda^2 U e^{2 \rho} \Phi
        +   F' \Theta,   \label{EQ2}  \\
    \Xi_{\pm} &=& 
           F' (\Theta \mp {\dot{\phi}_b}\partial_{\pm}\Phi)  \label{EQ3}.
\end{eqnarray}
Let's transform these equations into more tractable form.  
First, we solve $\Theta$ formally from the equations (\ref{EQ1}) and  
(\ref{EQ2}),  
\begin{eqnarray*}
    \Theta   &=&  
              - 2 \lambda^{2} e^{2 \rho_b}
               \frac{F'U - \frac{\kappa}{2} U'}{F'^2 - 4 \kappa G}
               \Phi.  
\end{eqnarray*}
Next, we define a new variable $\tilde{\Phi}$ by equation (\ref{phitilde}).
We note that from the equation (\ref{EQ3}) we can express $\Xi$ with 
$\tilde{\Phi}$,  
\begin{eqnarray*}
    \Xi_{\pm} &=&  
                \mp {\dot{\phi}} F' 
                 \exp \left[2\int \frac{\Omega}{\dot{{\phi}_b}} dr \right]
                 \partial_{\pm}\tilde{\Phi}. 
\end{eqnarray*}

Finally we obtain the equation (\ref{wave}) from the  
equation (\ref{phitilde}), (\ref{EQ1}) and (\ref{EQ2}),     

\begin{eqnarray}
    \left[  \frac{\partial^{2}}{\partial t^{2}} 
      - \frac{\partial^{2}}{\partial r^2}
      + V[\rho_{b}(r),\phi_{b}(r)] \right] \tilde{\Phi}(x)
     &=& 0, \nonumber 
\end{eqnarray} 
where
\begin{eqnarray}
V &=& 2 \frac{d}{dr} \left[\frac{\Omega}{\dot {\phi}_b} \right]
   +  4 \left[\frac{\Omega}{\dot{\phi}_b} \right]^2
   +  4 \frac{\dot{\rho}_b}{\dot{\phi}_b} \Omega
   +  \frac{16 G}{F'} \Omega
   +  4 \lambda^{2} e^{2 \rho_b} \frac{U'}{F'}  \label{potential} \\ 
\mbox{} &=&  
\frac{32\kappa \lambda^2 e^{2\rho_b}}
{{\dot{\phi}_b}^2 U ({F'}^2 -4\kappa G )^2 }
\left[
a{\dot{\phi}_b}^2
-2b{\dot{\phi}}_b {\dot{\rho}}_b
+c({\dot{\rho}_b}^2 -\lambda^2 )
\right] \label{POTinFGU},
\end{eqnarray}
with
\begin{eqnarray*}
a&=&
\frac{1}{2}GF''U^2 -\frac{1}{4}G'F'U^2
+4G^2 U^2-GF'UU'
+\frac{1}{16}({F'}^2 UU''-F'F''UU' )\nonumber\\
&&
+\frac{\kappa}{8}
(2G{U'}^2 -2GUU''+G'UU' ),\\
b&=&
\frac{1}{8}(F'U'-8GU)\left(F'U-\frac{\kappa}{2}U' \right),\\
c&=&
\frac{1}{4}\left(F'U-\frac{\kappa}{2}U' \right)^2.
\end{eqnarray*}
From the dependence of $\kappa$,    
we can realize that this potential is due to fully quantum effect. 

We can also express the potential in terms of $\tilde{F}$, (\ref{POTinF}),    
with the aid of the equation (\ref{Ftilde}), (\ref{GsEQ}), (\ref{UsEQ})
and the relation 
\begin{eqnarray*}
&&
\Omega = -\frac{\dot{\phi}_b}{2} 
          \frac{ \dot{\tilde{F}} \ddot{\tilde{F}} 
                -\frac{\kappa}{2} F^{3}}
               {\dot{\tilde{F}}^2 -
                \kappa\ddot{\tilde{F}} -
                \kappa^2 \lambda^2 }.
\end{eqnarray*}

\section{Potential to Leading Order of $\kappa$}
In this section we shall study the potential obtained in Appendix B 
to leading order of
$\kappa$ and derive the equations (\ref{perturbPOT}) and (\ref{POTinLarge}). 
We use the technique which is discussed in \cite{FHI}.
Now we write our variables as
\begin{eqnarray}
&&
\phi =\phi_{c} +\kappa \phi_{c1}, 
\nonumber\\
&&
\rho=\rho_{c} +\kappa \rho_{c1},
\nonumber\\
&&
F=F_c +\kappa F_{c1},
\nonumber\\
&&
\bar{G}=\bar{G}_c +\kappa \bar{G}_{c1},
\nonumber\\
&&
U=U_c +\kappa U_{c1},
\nonumber
\end{eqnarray}
where $\phi_{cn}$ denotes the $\phi$ to order $\kappa^{n}$ and so on.  

Then the potential takes the form
$$
V=V_{c1} + V_{c2} +\cdots.
$$
Note, the lowest order of potential is generally $\kappa^1$ 

To the leading order of $\kappa$, the equation (\ref{FsEQ}) is
$$
\frac{F^{(3)}_c}{\ddot{F}_c}
-2\frac{\ddot{F}_c}{\dot{F}_c}
=
\frac{\ddot{\rho}_c}{\dot{\rho}_c}-2\dot{\rho}_c.
$$
This can be solved formally as
\begin{eqnarray}
\dot{F}_c = \frac{2 e^{\rho_c(\infty)} \mu}{e^{2\rho_c(\infty)-2 \rho_c} - 1},
\nonumber
\end{eqnarray}
where $\mu$ is black hole mass. This is the equation (\ref{classicalF}). 

By substituting this into the equation (\ref{POTinF}),
we obtain the equation (\ref{perturbPOT})

Moreover we can make a next step to rewrite the potential for 
a black hole with larger mass $M$ than $\mu$.  
The gravity field of the black hole with larger mass which is denoted 
by $\rho'_c(r')$ can be expressed as
\begin{eqnarray}
&&
e^{2\rho'_c}
= \frac{e^{-2\rho_c(\infty)}}{B^2} 
  \left[ 1 - \frac{M}{\mu} (1-e^{2 \rho_c(r)-2\rho_c(\infty)}) \right].
\label{rho'rho}
\end{eqnarray} 
Here the relation between $r$ and $r'$ is  
\begin{eqnarray}
\frac{dr'}{dr}
=
\frac{\dot{\rho}_c(r_H)}{\lambda}  \frac{d\rho'_c}{d\rho_c}.
\label{r'r}
\end{eqnarray} 

The horizon in $r'$ coordinate is at $r_H$, defined by (\ref{rH}) in $r$ 
coordinate.

From the equation (\ref{perturbPOT}), (\ref{rho'rho}) and (\ref{r'r})
we obtain the equations (\ref{integratedr'inr}) and (\ref{POTinLarge}).

Finally, we note that the area of the potential to order of $\kappa$  can
be obtained analytically.  From equation (\ref{perturbPOT}), we can
obtain 
\begin{eqnarray}
\int_{-\infty}^{\infty} V_{c1}(r) dr  &=& \frac{\kappa \lambda^2}{\mu} 
        e^{\rho_c(\infty)} 
        \left[1 - e^{2\rho_c(\infty)}\frac{R_{H}}{4 \lambda^2}\right]
\end{eqnarray}
where, $R_H$ is the scalar curvature at horizon.
Thus we can see that there is a regime where the area is positive.

\section{Instability analysis}
In this appendix we apply the formalism discussed above to study 
the dynamical instability of black holes.

For CGHS model \cite{CGHS}, we take 
$$
F = G = U =  e^{-2 \phi}    
$$
and from the equation (\ref{POTinFGU}),  we obtain
\begin{eqnarray*}
V_{c1} &=&
        -  \frac{\kappa^2 \lambda^3}{\mu}
           \frac{ e^{-2\lambda r} }{(1+\frac{\mu}{\lambda}e^{-2\lambda r} )^3}
< 0,  
\end{eqnarray*}
where $\mu$ is the black hole mass.

For Lowe model \cite{L}, we take
$$
F = 2 G = e^{-2 \phi}  
$$
and
$$
U =1.
$$ 
In this case the classical solution is well-known and take the form: 
\begin{eqnarray*}
&&
e^{2\rho_c} = 1 - \frac{1}{2 \lambda \overline{r}},  \\
&&
e^{-2 \phi_c} = 2 {\lambda}^2 {\overline{r}}^2,  
\end{eqnarray*}
where 
\begin{eqnarray*}
&&
r = \overline{r} + \frac{1}{2 \lambda} \ln (2 \lambda \overline{r} -1). 
\end{eqnarray*}
Then from equation (\ref{perturbPOT}) we obtain 
\begin{eqnarray*}
V_{c1} &=&
        -  \frac{\kappa}{2 {\overline{r}}^5}
           (\overline{r} - {\overline{r}}_H)^2
           (\overline{r} + 3 {\overline{r}}_H)
< 0,  
\end{eqnarray*}
where 
\begin{eqnarray*}
{\overline{r}}_H = \frac{1}{2 \lambda} = 2 \frac{\mu}{m_{pl}^2}    
\end{eqnarray*}
and $m_{Pl}$ is the Planck mass.
It was shown numerically in \cite{IHMF} that the potential in both
model to full order of $\kappa$ is also negative,
thus the static quantum black hole solutions found in \cite{L, BGHS}
are unfortunately unstable against linear perturbation.

Next,we study the solvable model.
If we impose the following condition on $F, G$ and $U$  
\begin{eqnarray}
G=\frac{F'}{4}\frac{U'}{U} -\frac{\kappa}{16} 
\left[\frac{U'}{U} \right]^2,
\end{eqnarray}
then we can analytically  solve this type of model as can be seen 
in the following.  
For RST model \cite{RST} we take 
\begin{eqnarray*}
&&F = e^{-2\phi} -\frac{\kappa}{2}\phi,\\
&&G=U= e^{-2\phi} 
\end{eqnarray*}
and one may easily show that RST model is included in this type model.

In the solvable model, it is convenient to use variables:
\begin{eqnarray}
&&\chi=\rho +\frac{1}{2} \ln U,\\
&&\eta=F-\frac{\kappa}{2}\ln U.
\end{eqnarray}
With these variables, the equations of motion 
(\ref{motion1}), (\ref{motion2}) and
the constraint equation (\ref{constraint}) take the form:
\begin{eqnarray}
&&\partial_+ \partial_- \chi=0,\\
&&\partial_+ \partial_- \eta= -\lambda^2 e^{2\chi},\\
&&
\partial_{\pm}^2 \eta 
-2\partial_{\pm} \chi \partial_{\pm} \eta
=
-\kappa[\partial_{\pm}^2 \chi -(\partial_{\pm} \chi )^2 +t_{\pm} ].
\end{eqnarray}

For the static case, these are 
\begin{eqnarray}
&&\ddot{\chi_b}=0,\\
&&\ddot{\eta_b}=4\lambda^2 e^{2\chi_b},\\
&&\ddot{\eta_b} -2\dot{\chi_b}\dot{\eta_b}=
-\kappa [\ddot{\chi_b}-\dot{\chi_b}^2 + \lambda^2 ].
\end{eqnarray}
We can solve these analytically as
\begin{eqnarray}
&&\chi_b =\lambda r,\\
&&\eta_b = e^{2\lambda r} + const.
\end{eqnarray}
where boundary conditions are taken as (\ref{rhosBC}) and (\ref{FsBC}) 
and we choose $\rho_o = 0$.

According to section 4, defining  $\Phi$ and $\tilde{\Phi}$
which in this case become 
\begin{eqnarray*}
\Phi
&=&\delta\chi - \frac{\partial}{\partial r}
\left[\frac{\delta\eta}{\dot{\eta}_b} \right]
-\frac{d\chi}{d\eta}|_b \delta\eta, 
\\
\Phi&=&e^{-2\lambda r} \tilde{\Phi},
\end{eqnarray*}
then from (\ref{wave}) and (\ref{potential}) we obtain the equation of 
motion for $\tilde{\Phi}$ as
\begin{eqnarray}
\partial_+ \partial_- \tilde{\Phi} =0.
\end{eqnarray}
This equation says that the solvable models, including RST model, has no 
potential and thus are stable against the linear perturbations.

\
\\

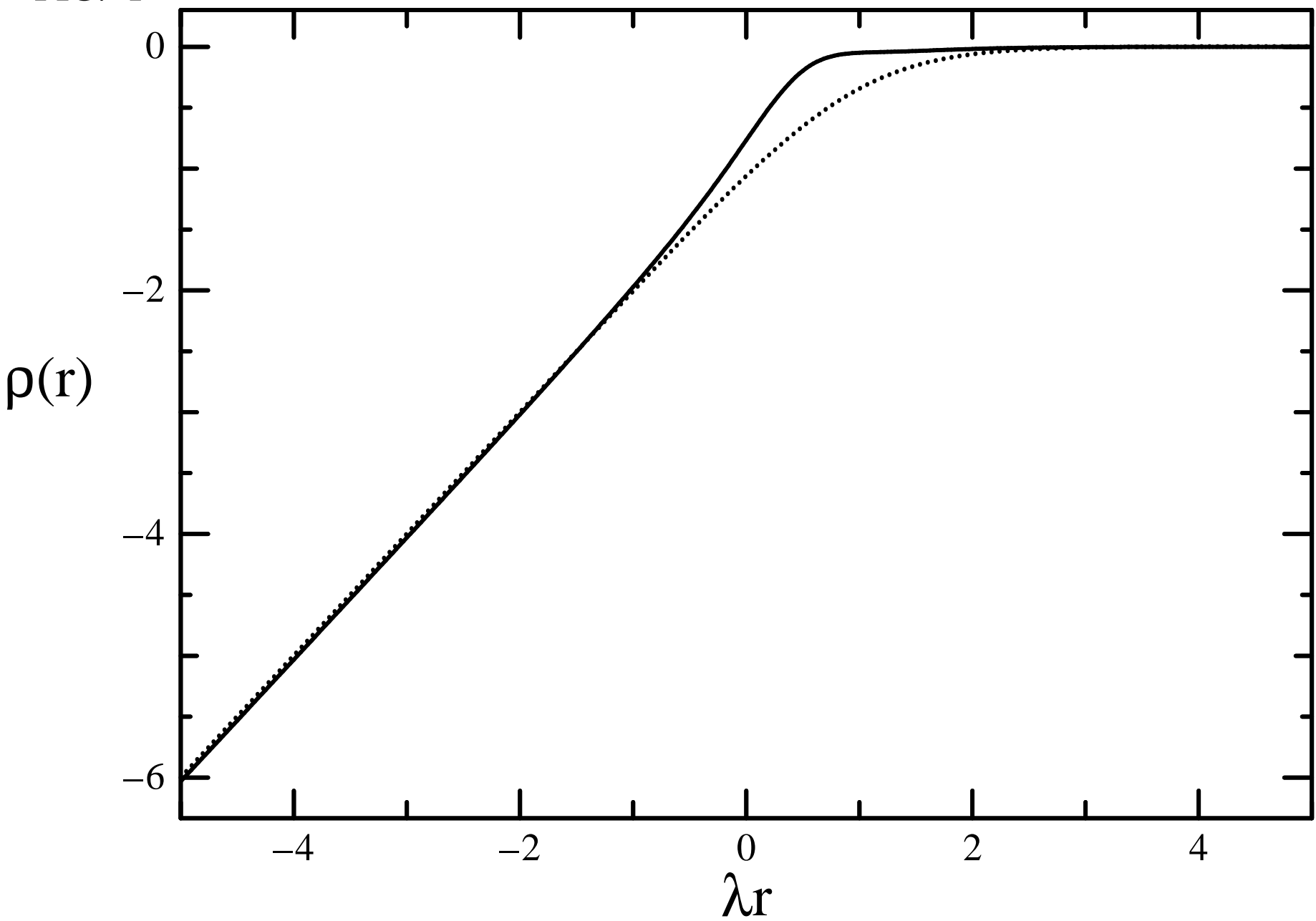
FIG. 1. 
A plot of gravity field $\rho(r)$  (solid line, see the equation 
(\ref{rho})) and 
 $\rho_{\rm CGHS}(r)$  (dashed line) vs $\lambda r$. Here,
$\rho_{CGHS}(r) = - \ln (1 + e^{-2 \lambda (r-1)})/2$ is the classical 
solution of the CGHS model with mass $\mu = \lambda$.

\ \\

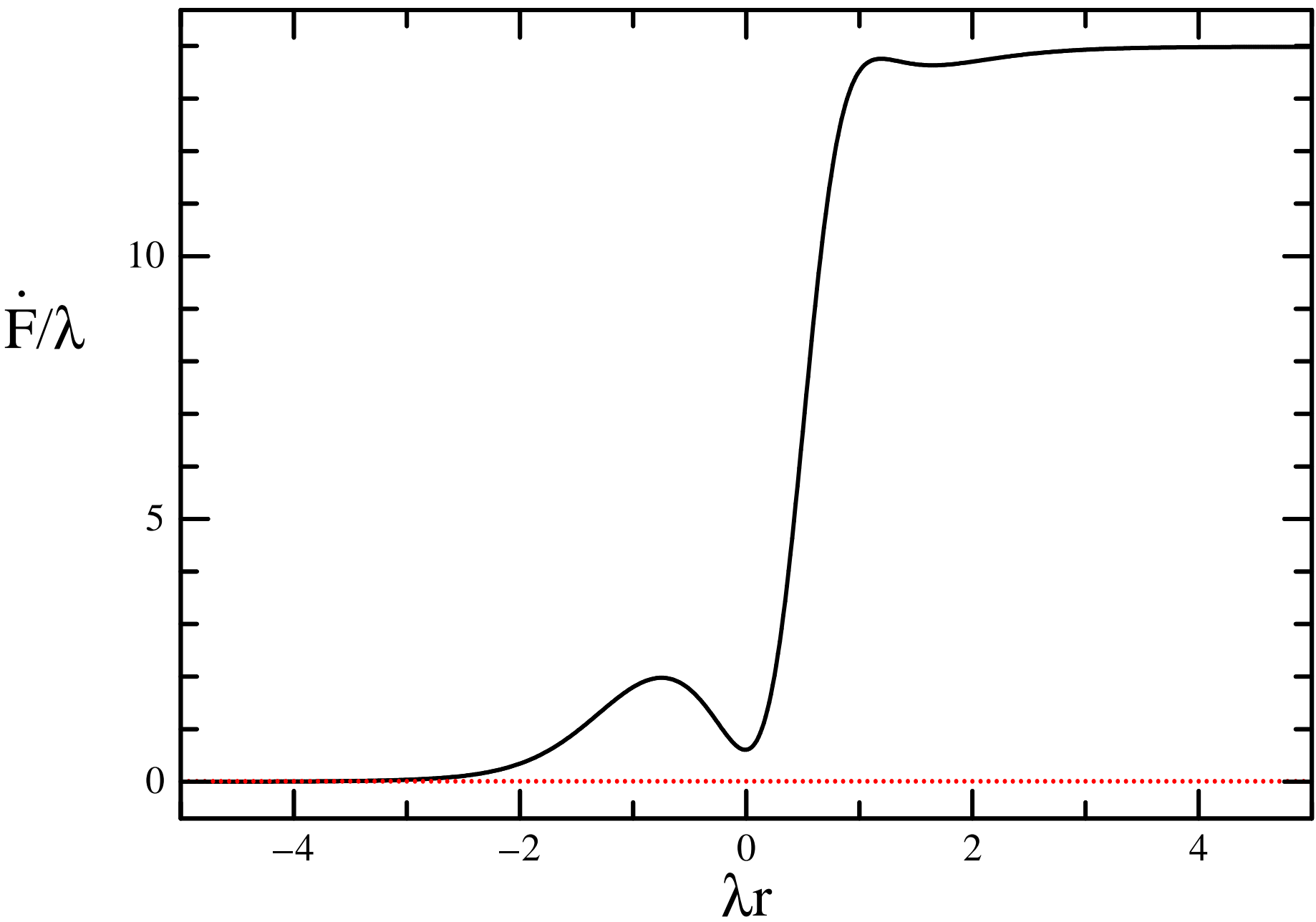
FIG. 2.
A plot of $\dot{F}/\lambda$ vs $\lambda r$. $\dot{F}$ is obtained
numerically from the equation (\ref{FsEQ}). Here, the 
gravity field is chosen as the equation (\ref{rho}). 

\ \\

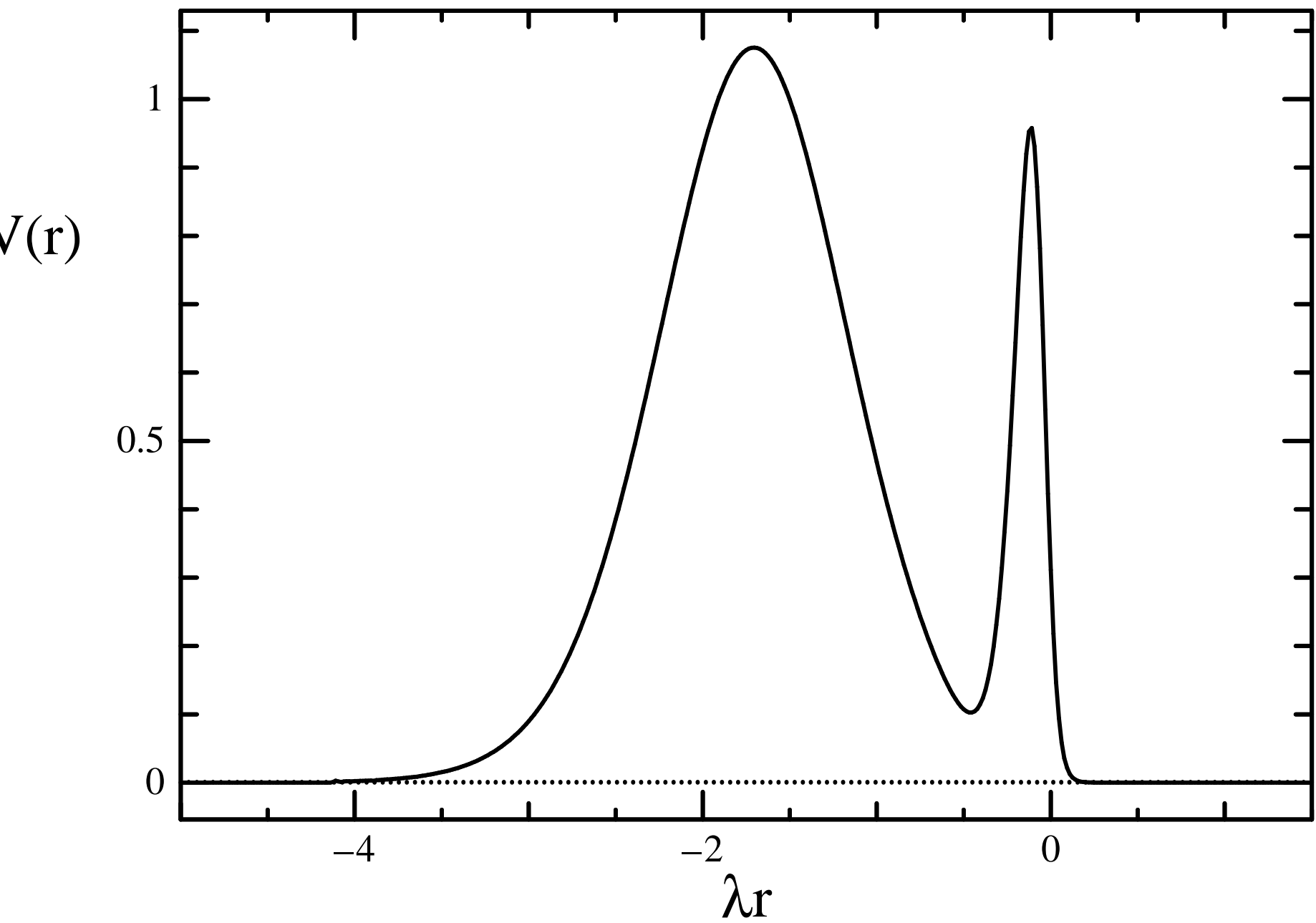
FIG. 3.
A plot of the potential $V(r)$ to full order of $\kappa$ vs $\lambda r$   
(see the equation (\ref{POTinF})). $V(r)$ is evaluated with 
$\dot{F}$ shown in Figure 2.

\ \\

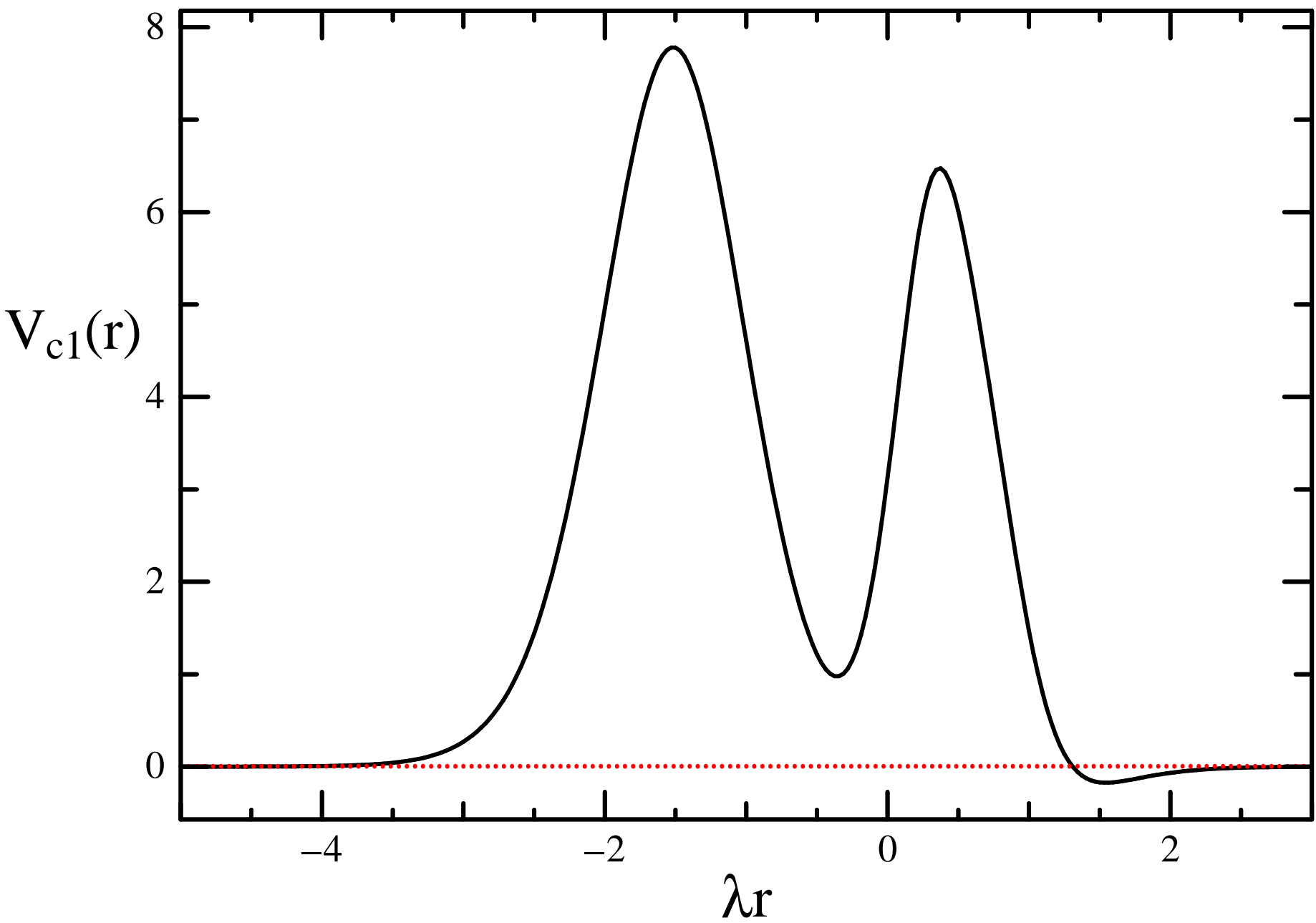
FIG. 4.
A plot of potential $V_{c1}(r)$ vs $\lambda r$. 
$V_{c1}(r)$ is calculated from the equations 
(\ref{perturbPOT}) and (\ref{rho}).

\ \\

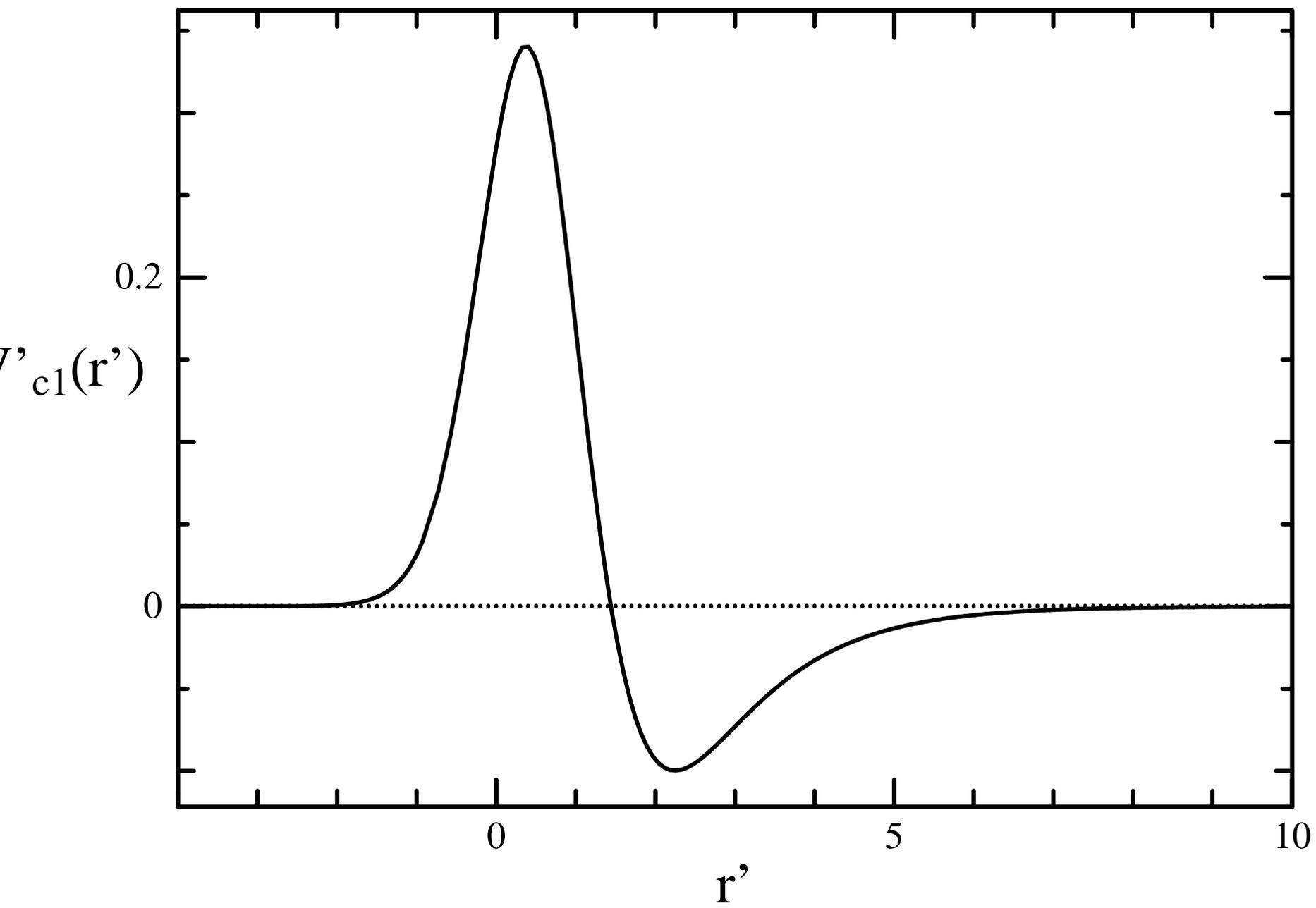
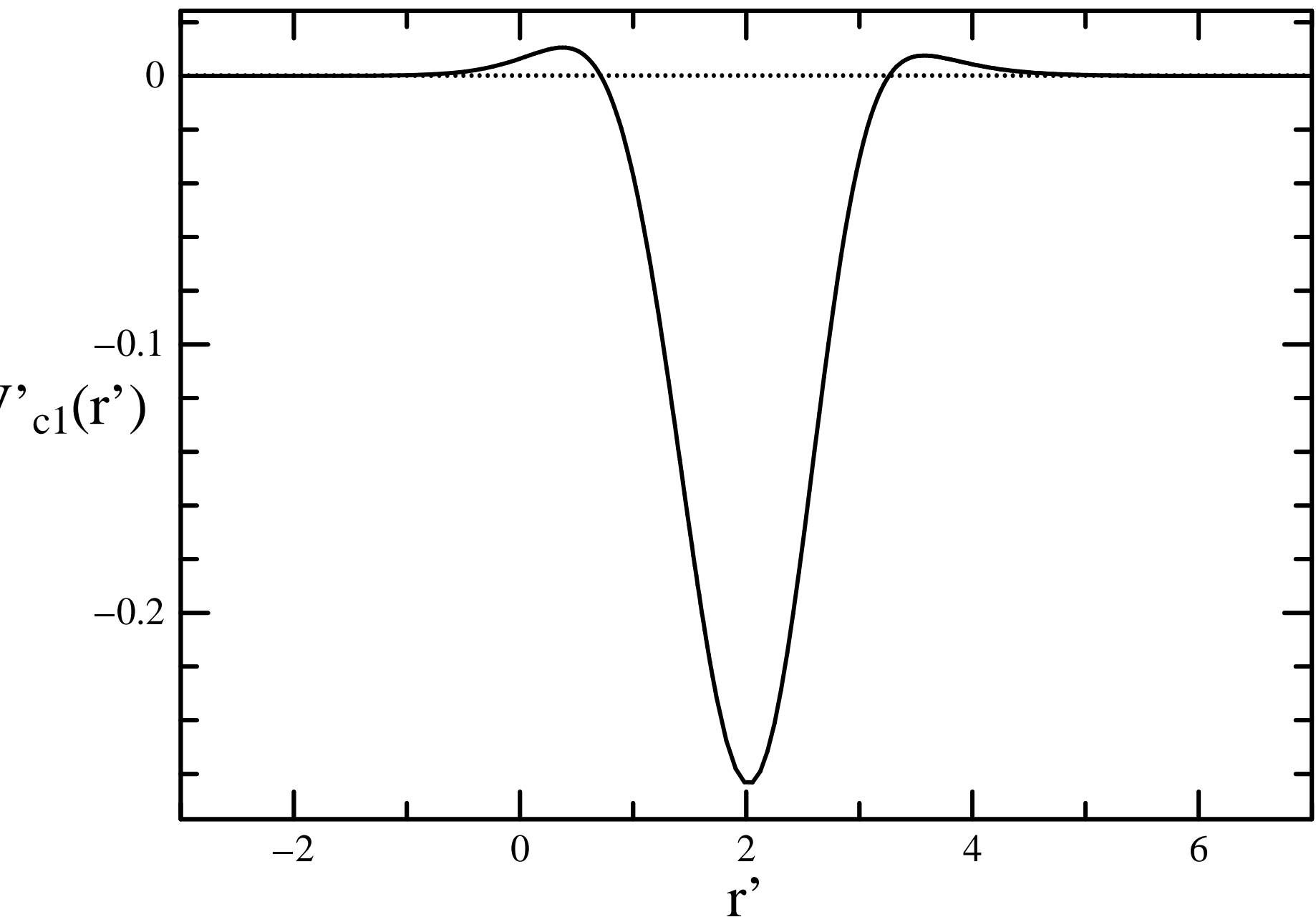
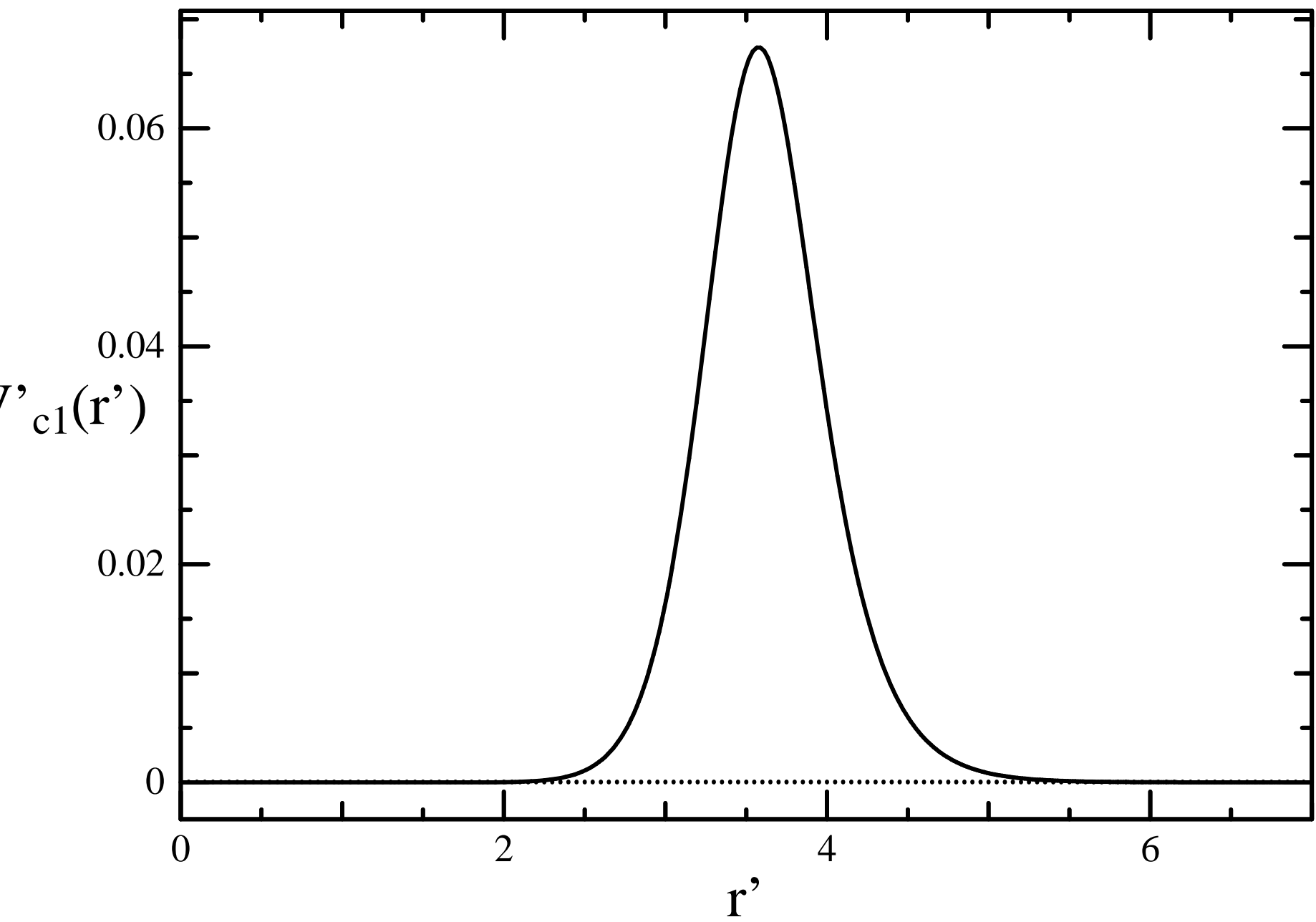
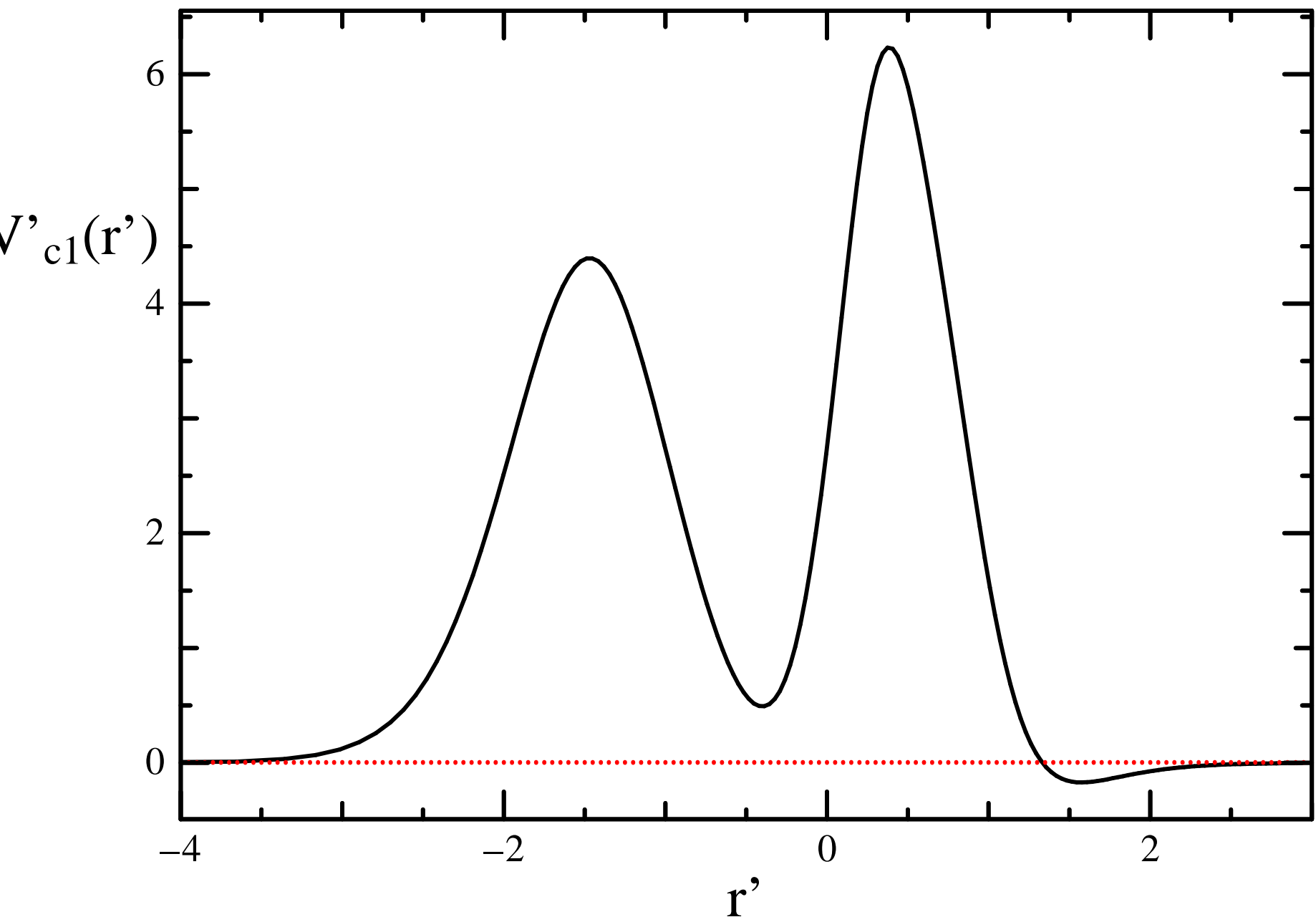
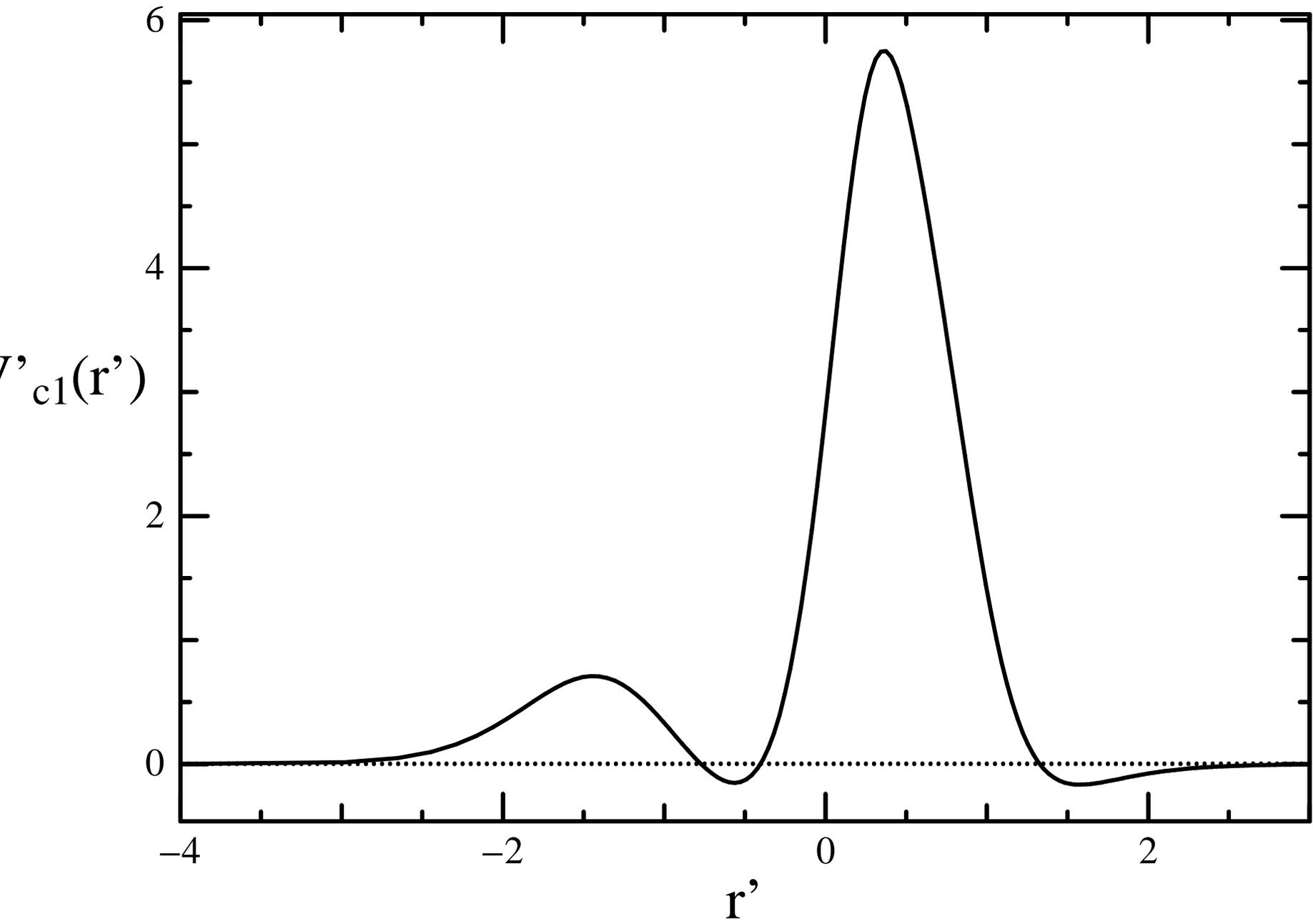
FIG. 5.
Plots of potentials to order $\kappa$ vs $r'$.
Note that the absolute value of $r'$ is 
meaningless in  FIG. 5 and 6 because there is 
ambiguity of an integral constant in the equation (\ref{integratedr'inr}).  
In FIG. 5, the cosmological term $\lambda$ retain its value 
when $M$ increase.
(a) A potential ${V'}_{c1}(r')$ calculated from the equations 
(\ref{POTinLarge}) and (\ref{rho}). Here, $M/\mu = 1.001$ ;       
(b) $M/\mu = 1.005$ ; (c) $M/\mu = 2$ ; (d) $M/\mu = 10$ ; (e) $M/\mu = 15$.
Note that, in some parameter $M/\mu$ space the potential is negative in 
almost region (foe example, see FIG. (c)). 
In such parameter space unstable modes exist and 
thus the coresponding static black hole is dynamically unstable
(see Appendix D).  We do not adopt a black hole with such a parameter.

\ \\

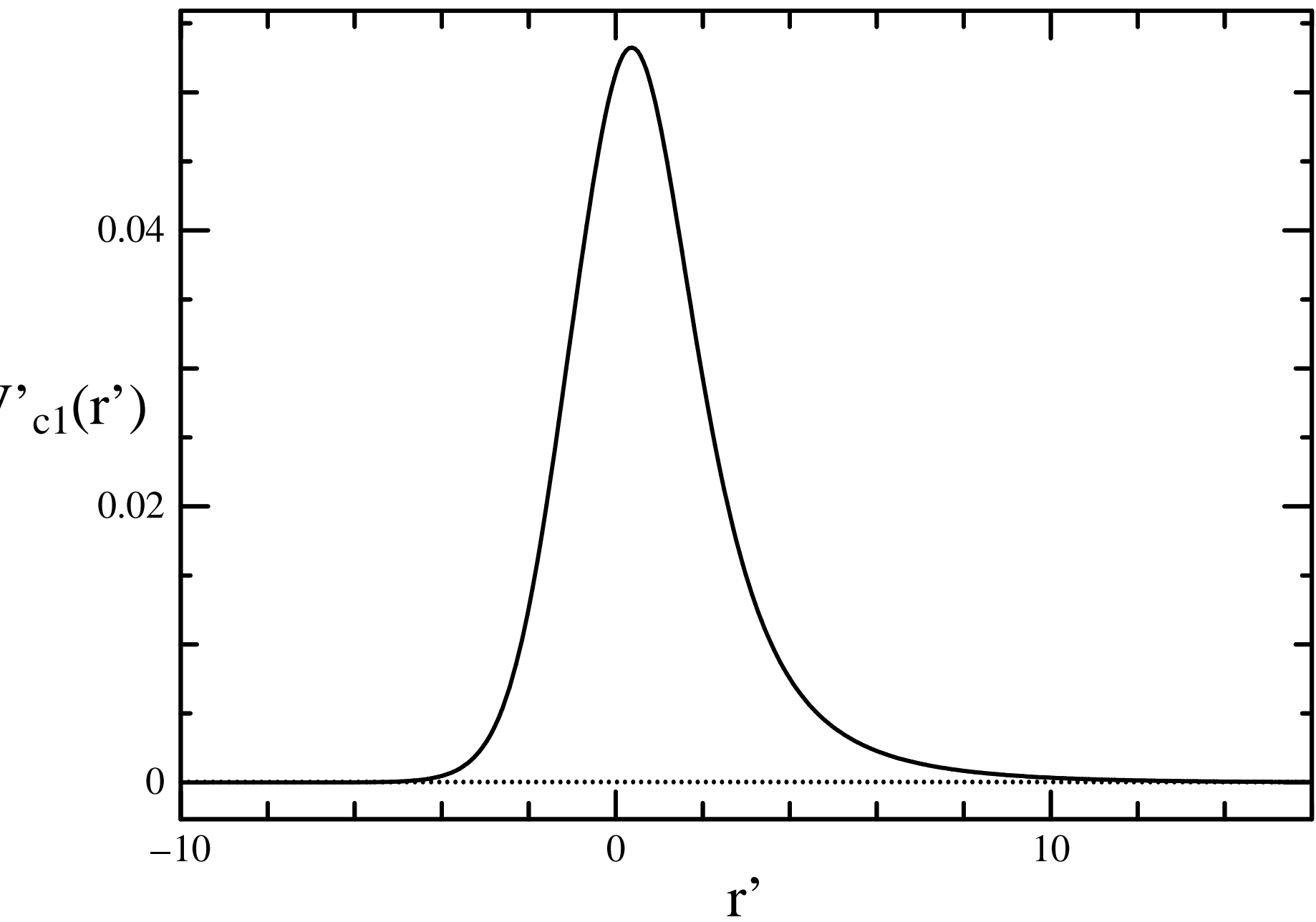
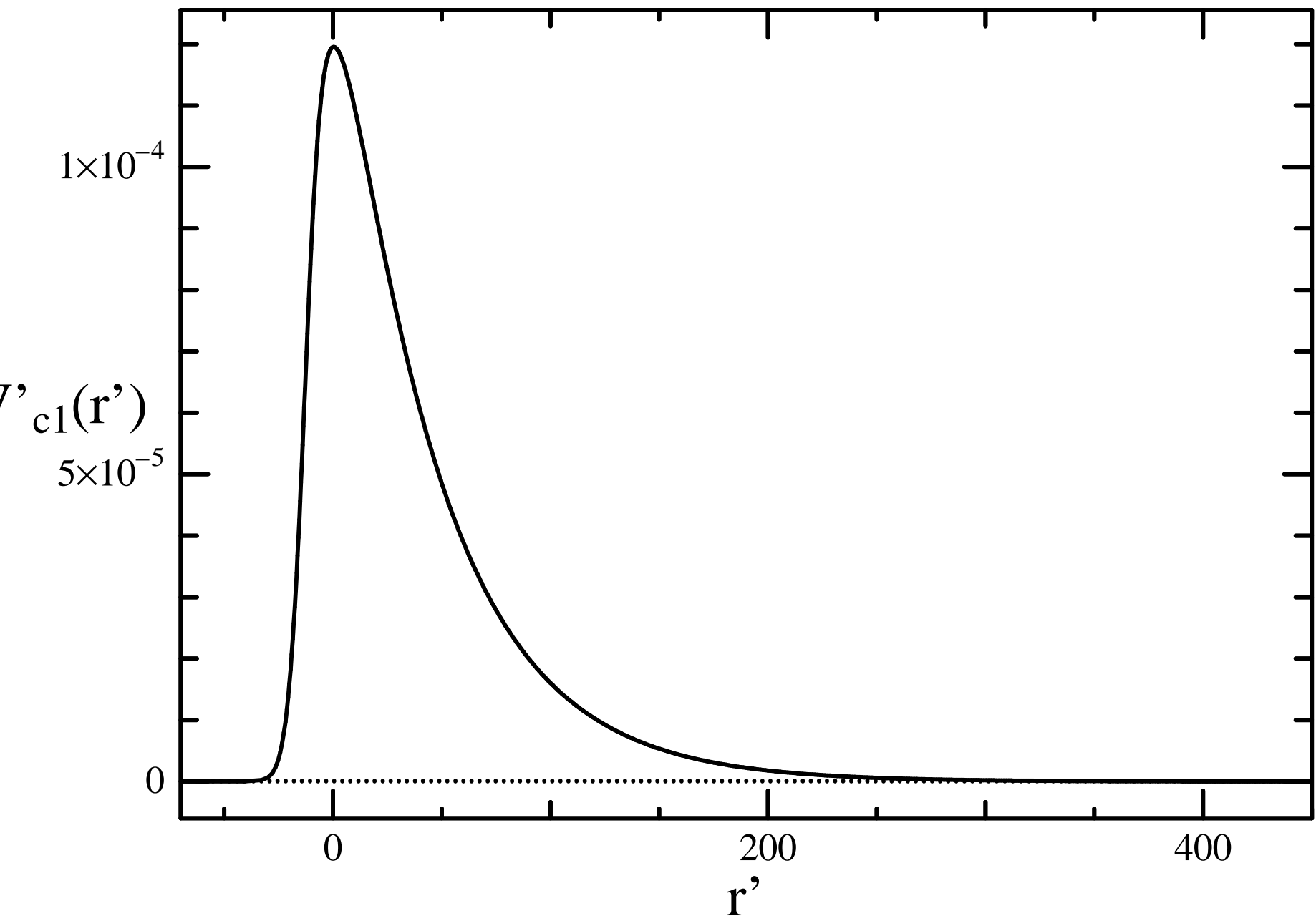
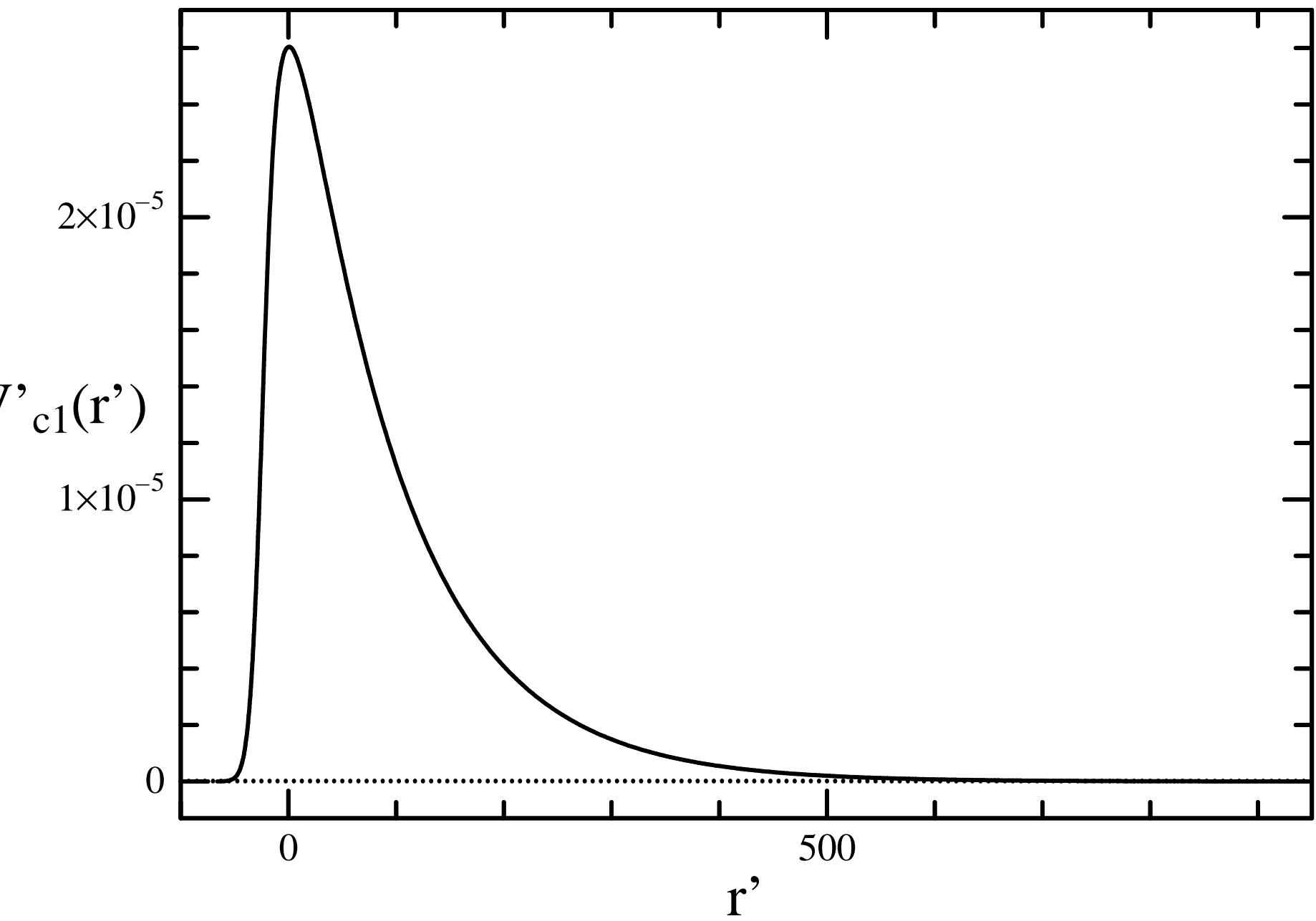
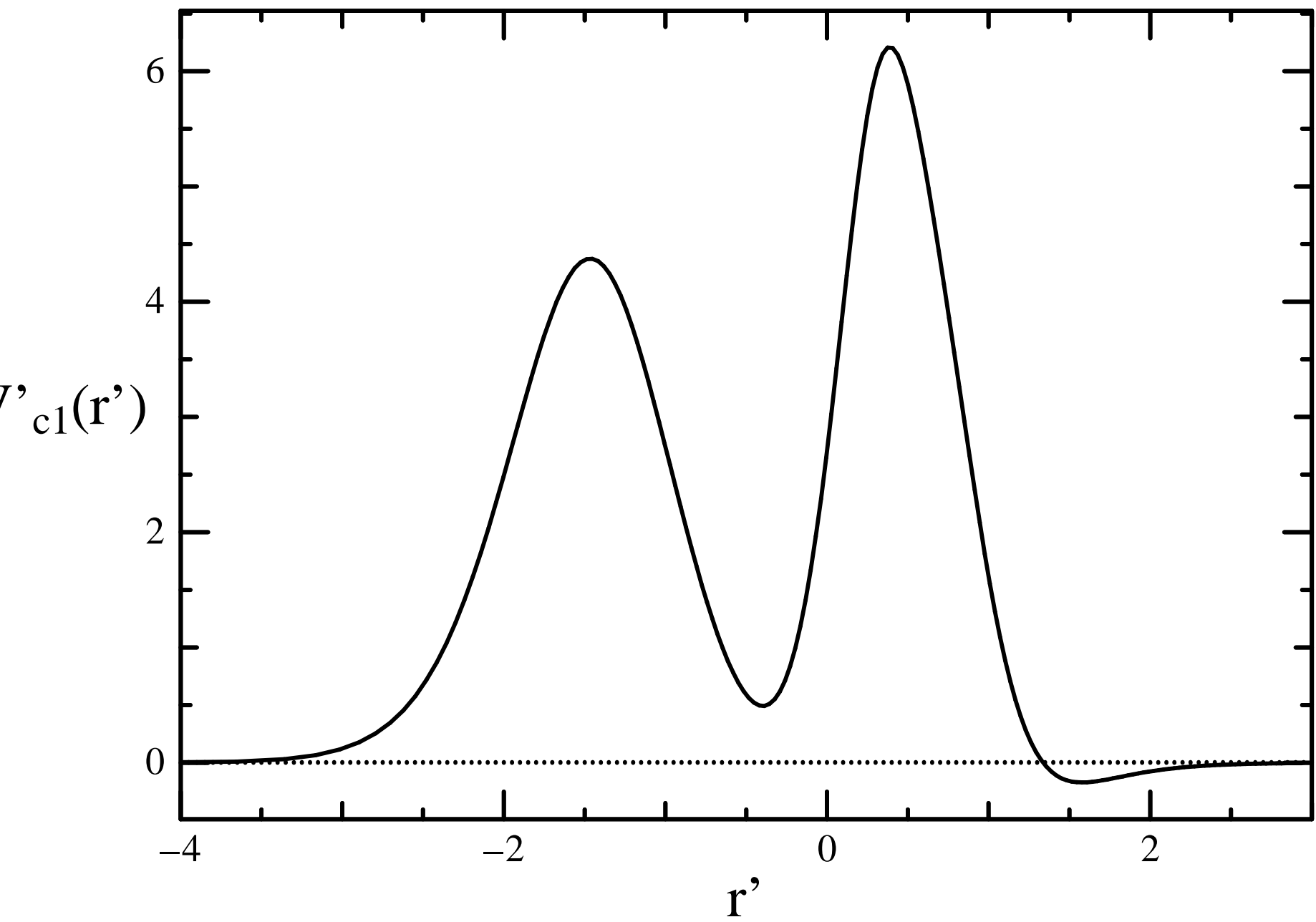
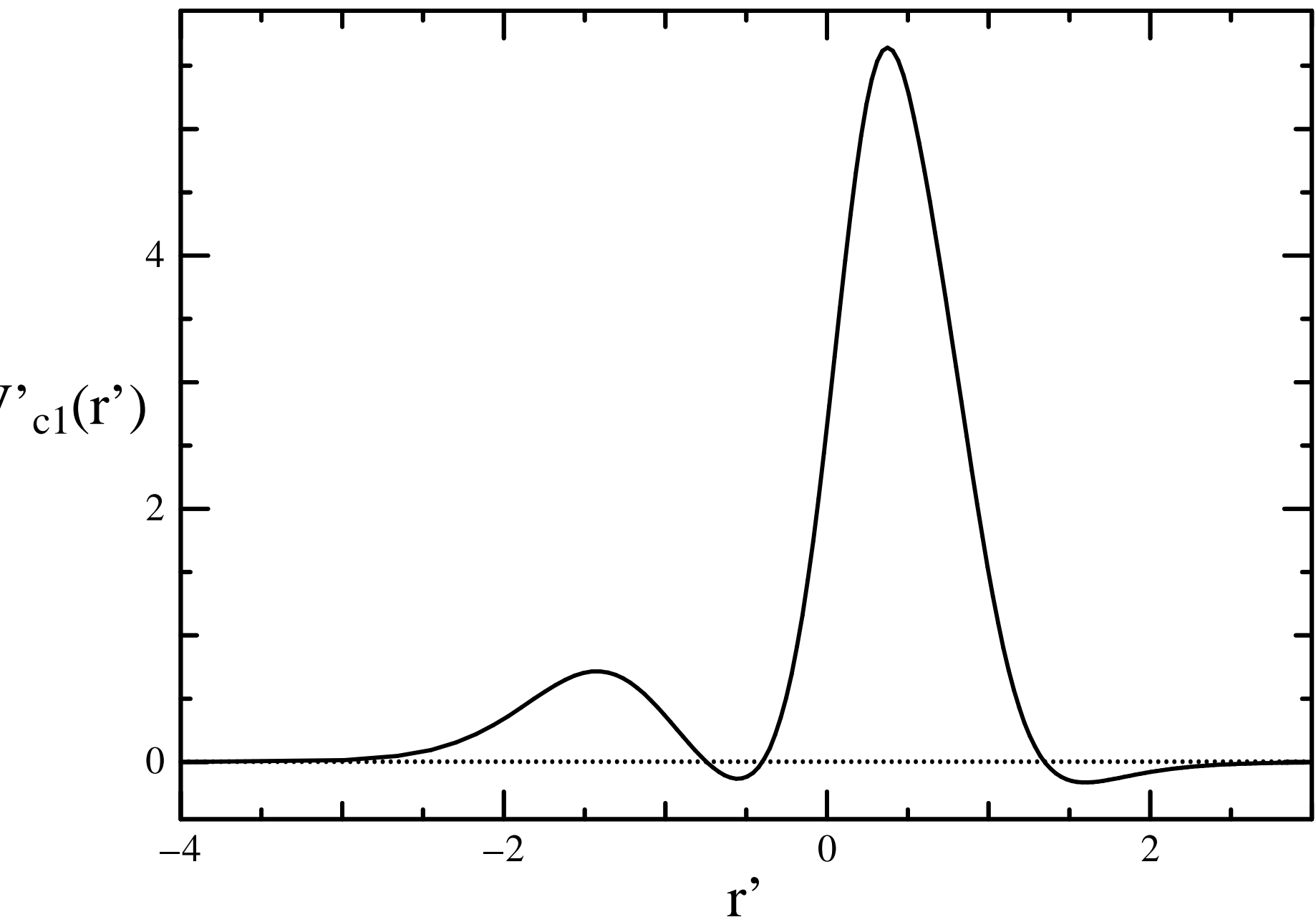
FIG. 6.
Plots of potentials to order $\kappa$ vs $r'$.
In FIG. 6, $\lambda$ scales as $M^{-1}$.
(a) A  potential ${V'}_{c1}(r')$ calculated from the equations 
(\ref{POTinLarge}) and (\ref{rho}). Here, $M/\mu = 1.001$ and 
$\lambda = 1/1.001$ ; (b) $M/\mu = 1.005$ and $\lambda = 1/1.005$ ; 
(c) $M/\mu = 2$ and $\lambda = 1/2$ ; 
(d) $M/\mu = 10$ and  $\lambda = 1/10$ ; 
(e) $M/\mu = 15$ and $\lambda = 1/15$.

\end{document}